\documentclass[12pt]{article}
\usepackage{amsmath,amssymb,amsthm,color}
\usepackage{graphicx}
\usepackage{cite}
\usepackage{epsfig}
\usepackage{lineno}
\renewcommand{\baselinestretch}{1.66}

\begin{document}
\title{Unusual Electronic excitations in ABA trilayer graphene \\}
\author{
\small Chiun-Yan Lin$^{a}$, Ming-Chieh Lin$^{b,*}$, Jhao-Ying Wu$^{c}$, Ming-Fa Lin$^{a,d,e*}$ $$\\
\small  $^a$Department of Physics, National Cheng Kung University, Taiwan\\
\small  $^b$Multidisciplinary Computational Laboratory, Department of Electrical and Biomedical \\
\small  Engineering, Hanyang University, Seoul 04763, South Korea\\
\small  $^c$Center of General Studies, National Kaohsiung Marine University, Kaohsiung, Taiwan\\
\small  $^d$Hierarchical Green-Energy Materials Research Center, National Cheng Kung University, Tainan, Taiwan\\
\small  $^e$Quantum topology center, National Cheng Kung University, Tainan, Taiwan\\
 }
\renewcommand{\baselinestretch}{1.66}
\maketitle

\renewcommand{\baselinestretch}{1.66}

\begin{abstract}
The tight-binding model is closely associated with the modified random-phase approximation to thoroughly explore the electron-electron interactions in trilayer AB-stacked graphene. The intralayer and interlayer atomic/Coulomb interactions dominate the collective and electron-hole excitations. The unusual energy bands are directly reflected in the diverse transferred momentum-frequency phase diagrams. There exist three kinds of plasmon modes during the variation of the doping level, being accompanied with the complicated intraband and interband single-particle excitations.   The excitation behaviors are greatly diversified by the number of layers. The theoretical predictions require the high-resolution experimental examinations.
\end{abstract}

\par\noindent  * Corresponding author.
\\~{{\it E-mail addresses}: mflin@mail.ncku.edu.tw (M.F. Lin)}
\\~{{\it E-mail addresses}: mclin@hanyang.ac.kr  (M.C. Lin)}

\pagebreak
\renewcommand{\baselinestretch}{2}
\newpage

\vskip 0.6 truecm

The electron-electron Coulomb interactions are one of the main-stream topis in condensed-matter systems, especial for the emergent 2D materials with the intralayer and interlayer Coulomb/atomic interactions. The layered/few-layer graphenes are typical systems, since they possess the high-symmetry honeycomb lattice, distinct stacking configurations, tunable layer numbers, and weak but significant van der Waals interactions. The AB, AA, ABC, AAB stackings\cite{JCP129;234709,JAP109;093523,SurfSci610;53}, and the various twisted and sliding structures \cite{PRL109;126801} have been successfully synthesized in experimental laboratories. Moreover, the nature graphite mostly consists of the AB-stacked configuration\cite{PRSLSA181;101,PRSLSA106;749}, and only part of system belongs to the ABC-stacked one\cite{PRSLSA106;749}. Apparently, the stacking configuration and layer number are shown to dominate the low-energy essential properties, e.g., energy bands and density of states (DOS) near the Fermi level ($E_F$), Their critical roles on the rich and unique electronic excitations in the AB-stacked trilayer graphene are the focuses of this current work. Comparisons with monolayer and bilayer systems are also made. Specifically, the modified random-phase approximation (RPA) is further developed to agree with the layer-dependent Coulomb interactions in 2D materials.

The Coulomb excitation behaviors are mainly determined by the electronic structures. The ${2p_z}$ orbitals of carbon atoms in graphene-related systems, which built the $\pi$ valence bands and the $\pi^\star$ conduction bands, are responsible for the electronic excitations lower than the middle frequency (${\sim\,6-10}$ eV). The low-lying band structures of layered graphenes strongly depend on the intralayer and interlayer hopping intergrals of ${2p_z}$ orbitals. The trilayer ABA stacking (Fig. 1(a)) exhibits the unusual energy bands, two pairs of parabolic valence and conduction bands and one pair of distorted Dirac-cone structures (Fig. 1(b)). The latter, as verified by the high-resolution angle-resolved photo emission spectroscopy (ARPES)\cite{PRL98;206802,Science313;951}, could survive in AB-stacked systems with odd layers\cite{PRL98;206802}. Also, there exist the special wavefunctions arising from the specific superpositions of the six tight-binding functions, being directly reflected in the existence/strength of the Coulomb interactions.  Band structure and electronic wavefunctions are thoroughly included in the current calculations.

There are a lot of theoretical predictions\cite{PRB87;235418,PRL81;4216,
PRB34;979,RevModPhys83;1193,PRB74;085406,ACSnano5;1026,PRB84;115420,
NanoLett16;6844,JPSJ75;074716,NJP8;318} and experimental measurements \cite{PRL105;016801,NanoLett14;3827,PRL83;161403,PRL80;113410,PRB61;7517} on electronic excitation spectra of layered graphenes.  Monolayer graphene, with the linear Dirac-cone structure, is predicted to show the intraband and interband electron-hole (e-h) excitations, and the low-frequency plasmon ($<2$ eV) under the extrinsic electron/hole doping\cite{PRB34;979,RevModPhys83;1193} and the temperature effect\cite{PRB87;235418} . The 2D acoustic plasmon, accompanied with the higher-frequency optical plasmon, could survive in bilayer AB stacking\cite{PRL81;4216}. However, the theoretical calculations have ignored some significant interlayer atomic interactions. They are absent for AB-stacked systems with layer number higher than 3. On the experimental side, the high-resolution electron-energy-loss spectroscopy (EELS)\cite{PRL83;161403,PRL80;113410} is successful used to verify/identify the electronic excitations in layered graphenes. The acoustic, $\pi$ plasmons and $\pi$ +$\sigma$ plasmons are examined to occur at the low ($\sim{0.1-1.0}$ eV)\cite{PRL83;161403}, middle ($\sim{5-7}$ eV) and high frequencies (${>14}$ eV)\cite{PRL80;113410}.  Another powerful equipment, that the inelastic light scattering spectroscopy has confirmed the low-frequency Coulomb excitations in doped semiconductors, e.g., ${\sim\,0.01-0.1}$eV plasmons\cite{PRB61;7517}.

In this work, the tight-binding model and the modified RPA are directly combined to thoroughly explore the diverse electronic excitations in trilayer ABA stacking. All the important intrinsic interactions are covered in the calculations. The dependence on the doping level is investigated in detail, being very useful in comprehending the significant effects due to the variation of the Fermi-Dirac distribution. The pristine system will be studied whether the interlayer-hopping-induced free electrons and holes could create the low-frequency acoustic plasmon. The dramatic changes in collective and e-h excitations (the transferred momentum-frequency phase diagram) are expected to be easily observed  as the doping level varies. How many kinds of plasmon modes and the diverse single-particle excitation regions is clearly identified by the delicate analyses. The predicted results could be verified by the experimental measurements.

The AB-stacked trilayer configuration is clearly shown in Fig. 1(a), in which two neighboring layers shift relative to each other by one C-C bond length ($b$) along the armchair direction ($\hat x$). There are six carbon atoms in a primitive unit cell. The contributions due to the ${2p_z}$ orbitals of carbon atoms are sufficient for the low-energy energy bands and  electronic excitations. The zero-field Hamiltonian is built from the six tight-binding functions associated with the (${(A^1,B^1,A^2,B^2, A^3,B^3)}$ sublattices. The superscript $i$ represents the $i$-th layer.
The ${6\times6}$ Hermitian matrix covers the non-vanishing elements related to the nearest-neighbor intralayer hopping integral (${\gamma_{0}=-3.12}$ eV), three neighboring-layer hopping integrals ($\gamma_1$ = 0.38 eV, $\gamma_3$ = 0.28 eV; $\gamma_4$=0.12 eV), two next-neighboring-layer hopping integrals ($\gamma_2$=$-$0.021 eV; $\gamma_5$=$-$0.003 eV), and the chemical environment difference between A and B sublattices ($\gamma_6$=$-$0.0366 eV). The details of the Hamiltonian matrix could be found in Ref.\cite{PRB43;4579}. All the significant atomic interactions are included in the tight-binding model.

The trilayer AB stacking has three pairs of low-lying valence and conduction bands, with a small band overlap\cite{IOPBook;978}. Few free carrier density in this semi-metallic pristine system will determine whether the acoustic plasmon could survive. One separated and distorted Dirac-cone structure (the first pair ($v_1$,$c_1$)) comes to exist near the Fermi level.  Their wavefunctions mainly come from the first and third layers, so no contributions from the second layer will be reflected in the bare response function (discussed latter in Fig. 3). Another two pairs of parabolic bands, respectively, appear roughly at ${E_F}$ (the second pair ($v_2$,$c_2$)) and $\pm\gamma_{1}$ (the third pair ($v_3$,$c_3$)). The valence bands are somewhat asymmetric to the conduction ones about ${E=0}$, while this property hardly affects the main features of the Coulomb interactions. That is, electron and hole dopings almost lead to the similar excitation behaviors. The former doping case is chosen for a model study, in which the Fermi level is located at the conduction bands.

The external Coulomb potential due to the incident electron beam is assumed to be uniform on each graphene layer. The $\pi$ and $\pi^\star$ electrons on the distinct layers will effectively screen the similar bare Coulomb potentials, leading to the charge redistributions and the induced Coulomb potentials. Moreover, the transferred momentum and frequency ($q,\phi,\omega$) are conserved during the dynamic electron-electron interactions, where $0^{\circ}\le\phi\le30^{\circ}$ is the angle between ${\bf q}$ and $\Gamma$K (the first Brilluoin zone in Fig. 1). By the Dyson equation, the effective Coulomb potential between two electrons on the $l$- and $l^\prime$-th layers is given by\cite{PRB74;085406}

\begin{equation}
\epsilon_{0}V^{eff}_{ll^{\prime}}(\mathbf{q},\omega)=V_{ll^{\prime}}(\mathbf{q})+
\sum\limits_{mm^{\prime}}V_{lm}(\mathbf{q})P^{(1)}_{mm^{\prime}}(\mathbf{q},\omega)
V^{eff}_{m^{\prime}l^{\prime}}(\mathbf{q},\omega)\text{.}
\end{equation}%

The first term is the bare Coulomb potential $V_{ll}^{\prime}=v_{q}e^{-q|l-l^{\prime}|}$ (the 2D potential ${v_q=2\pi\,e^2/q}$ associated with a 2D electron gas). The second term corresponds to the induced potential, in which the induced potential is proportional to the screening charge density using the Fourier-transform Poisson equation, and the latter is proportional to the effective potential under the linear self-consistent method. The linear coefficient, the bare polarization, which includes the layer-dependent electron-hole excitations, is expressed by

\begin{equation}
\begin{array}{l}
P_{mm^{\prime}}(\mathbf{q},\omega)=2\sum\limits_{k}\sum\limits_{h,h^{\prime}=c,v}\sum\limits_{n,n^{\prime}}
\biggl(\sum\limits_{s}
U_{smh}(\mathbf{k})
U^{\star}_{sm^{\prime}h^{\prime}}(\mathbf{k+q})\biggr)\\
\times\biggl(\sum\limits_{s}
U^{\star}_{smh}(\mathbf{k})
U_{sm^{\prime}h^{\prime}}
(\mathbf{k+q})\biggr)\times\frac{f(E^{h}_{n}(\mathbf{k}))-f(E^{h^{\prime}}_{n^{\prime}}(\mathbf{k+q}))}
{E^{h}_{n}(\mathbf{k})-E^{h^{\prime}}_{n^{\prime}}(\mathbf{k+q})
+\hbar\omega+i\Gamma}
\text{.}
\end{array}
\end{equation}%

Specifically, the excited electron and hole in each excitation pair, which arises from the Coulomb perturbation, frequently appear on distinct layers. $U_{smh}$'s are the amplitudes related to the six tight-binding functions, $s$ represents the specific A$^i$/B$^i$ sublattice, and $h$ denotes the valence/conduction state. Band-structure effects, using the layer-decomposed contributions, have been included in Eq. (2). From the detailed derivations under the Born approximation, the dimensionless energy loss function, being directly proportional to the measured EELS intensity, is defined as

\begin{equation}
\begin{split}
\mathbf{Im}[-1/\epsilon]&\equiv\sum\limits_{l}\mathbf{Im}\biggl[-V_{ll}(\mathbf{q},\omega) \biggr]
/\biggl(\sum\limits_{lm}V_{lm}(q)/N\biggl)\text{.}
\end{split}
\end{equation}%

The denominator is the average of all the external potentials on the $N$-layer graphene. Equation (3) is suitable for any emergent layered systems, such as, the group-IV and group-V 2D materials. This screened response function provides the full information on the diverse plasmon modes, and the bare one in Eq. (2) describes the single-particle (electron-hole) excitations.

The single-particle and collective excitations are dominated by energy bands and wave functions, being sensitive to the doping level. Electrons are excited from the occupied states to the unoccupied ones under the Fermi-Dirac distribution and the conservation of ${(q,\phi\,\omega)}$. The bare response function, corresponding to the e-h excitations, exhibits the special structure at the specific frequency, if the initial/final state of the allowed transitions comes from the band-edge state with the van Hove singularity, or the Fermi-momentum state with the step distribution. It should be noticed that some excitation channels are forbidden because of the symmetric/anti-symmetric properties of wave functions. The layer-dependent response functions consist of  four independent components: ${P_{11}}$=${P_{33}}$, ${P_{22}}$,  ${P_{12}}$ =${P_{21}}$=${P_{23}}$=${P_{32}}$; ${P_{13}}$=${P_{31}}$.

As for a pristine system, the available excitations include (${v_1\rightarrow c_1}$,${v_2\rightarrow c_2}$), (${v_1\rightarrow c_3}$,${v_3\rightarrow c_1}$); ${v_3\rightarrow c_3}$. in which the special structures, respectively, appear at very low frequency, $\sim 0.56$ eV and $\sim 0.92$ eV for $q=0.005 1/{\AA}$ and $\phi\,=0^\circ$. The imaginary parts of $P_{ll^{\prime}}$, as shown by the red curves, directly reflect the features of DOS and wave functions, and its special structure relies on the former. The obvious shoulder structures are due to the extreme states (the local maxima/minima). As a result, the symmetric peaks in the logarithmic forms are revealed in the real parts of $P_{ll^{\prime}}$ (black curves) by the Kramers-Kronig relations. Why the special structures at ${\omega\sim0.56}$ eV strongly rely on the distorted Dirac-cone bands ($v_1$ and $c_{1}$ bands) near ${E_F=0}$, since they are absent in $P_{21}$ closely related to the significant contribution of the second layer.

Electronic excitations are dramatically changed during the variation of the Fermi level. Part of them from the valence to conduction bands are suppressed by the electron doping, mainly owing to the drastic changes in the Fermi-Dirac distribution. However, there are more free carriers in conduction bands which could built the Fermi surfaces. In addition to the band-edge states, the Fermi-momentum ones, being closely related to the step distribution functions, create the special structures in the bare response functions. Only the pristine ${v_3\to\,c_3}$ interband excitations are independent of the electron doping, if the Fermi level is below the third conduction band, i.e., the special structure above 1 eV remains the similar form. For a ${E_F=0.2}$ eV system, conduction electrons will suppress three valence$\to$conduction excitations (${v_1\to\,c_1}$,${v_2\to\,c_2}$,${v_3\to\,c_1}$), and only the ${v_1\to\,c_3}$ could survive, leading to the special structure at ${\sim\,0.56}$ eV.  The lower-frequency special structures are generated by the Fermi surfaces. Most important, free carriers in conduction bands induce new ${c^i\to\,c^j}$ excitation channels, covering the intraband and interband transitions  simultaneously.  Both ${c_1\to\,c_1}$ and ${c_2\to\,c_2}$ intraband excitations could create the strong responses at the almost same low frequency ($<$0.1 eV), as indicated by blue arrows. Furthermore, the interband excitations, ${c_2\to\,c_1}$, ${c_1\to\,c_3}$ $\&$ ${c_2\to\,c_3}$, respectively, exhibit the special structures near ${0.32}$ eV, ${0.42}$ eV $\&$ ${0.5}$ eV. It should be noticed that the square-root divergent structures are frequently revealed in the imaginary and real parts of ${P_{ll^\prime}}$ because of the linear excitation energies and the Fermi-Dirac step function\cite{PRB34;979}. Apparently, the bare response functions will change in the further increase of $E_F$ and the variation of the transferred momentum (${(q\,,\phi\,})$; not shown).
The energy loss functions, the screened response spectra, are useful in understanding the plasmon modes and the Landau dampings. Furthermore, they directly correspond to the measured excitation spectra. The dimensionless Im${[-1/\epsilon\,]}$, as clearly shown in Figs. 4(a)-4(f), strongly depends on the  doping level and the magnitude ($q$) of the transferred momentum, but not $\phi$.  For a pristine system, it is difficult to observe the prominent peak in the loss spectra (the intensity lower than 0.2 in Fig. 4(a) at ${q=0.005}$ 1/$\AA$), indicating the collective excitations fully suppressed by the interband e-h excitations. Too few free carriers are responsible for the absence of the strong plasmon modes. Conduction electrons under doping can create two/one prominent peaks in excitation spectra (Figs. 4(b)-4(f) and insets), being identified from as the collective excitations. The lower-frequency plasmon. the first collective mode, has a rather strong intensity, since it is due to the intraband excitations of all the conduction carriers. However, the intensity of the higher-frequency plasmon is weaker but easily   observable for the sufficiently high $E_F$ (${E_F\ge\,0.4}$ eV). The second plasmon mode might arise from the ${c_{2}\to\,c_{3}}$ interband excitations. The energy loss spectra hardly depend on the direction of the transferred momentum, i.e., they are almost isotropic (Figs. 4(b) and 4(e)). Coulomb excitations are very sensitive to the magnitude. The plasmon frequencies grow with the increment of $q$, since the e-h excitation energies behave so, e.g., those of the first and second plasmons at different $q'$s in Figs. 4(b) and 4(f).

The (${\bf q\,,\omega}$)-phase diagrams could provide the full information on the single-particle and collective excitations, as clearly shown in Figs. 5(a)-5(f).  Any systems exhibit the vacuum regions which any excitations cannot survive, since electronic states of energy bands (Fig. 1) do not create some (${\bf q\,,\omega}$) Coulomb interactions. For a pristine system, there are no obvious plasmon modes, according to the EELS intensities in the whole (${\bf q\,,\omega}$) range (Fig. 5(a)). The boundaries of the ${c_i\to\,v_j}$ interband excitations are characterized by the band-edge states at the K/K$^\prime$ point. The e-h Landau dampings are very strong and effectively suppress the plasmon modes. After the electron/hole doping, all the e-h excitation boundaries are dramatically altered by the distinct Fermi surfaces (Figs. 5(b)-5(f)) except that the highest-frequency ${v_3\to\,c_3}$ might remain similar under $E_F\le\,E^{c_{3}}(K)$. Apparently, the single-particle excitations are enriched by the new ${c_i\to\,c_j}$ excitation channels (the solid and dashed curves). Each doped system could display the strongest acoustic plasmon, with the ${\sqrt q}$-dependent frequency at long wavelength limit, as previously verified in a 2D electron gas\cite{PRB34;979}. The first plasmon mode will gradually decay in the increment of $q$ and disappear at the critical momentum. $q_c$ grows with the increasing doping level (Figs. 5(b)-5(e)). The second plasmon related to the ${c_2\to\,c_3}$ excitations is identified as a optical mode because of a finite frequency at ${q\to\,0}$. Its frequency and $q$ do not have a simple relation. The spectral intensity first increases, reaches the maximum, and then declines. This plasmon is easily to be observed with the increasing doping level (e.g., 0.8 eV in Fig. 5(f)). Moreover, there exists the third plasmon in between the first and second modes under the Fermi level through the $c_3$ band. For example, it is revealed in ${E_F}$=0.6 eV and 0.8 eV (Figs. 5(d) and 5(e)). Its intensity is lowest among three plasmon modes. The third mode is examined to come from the ${c_1\to\,c_3}$ excitations, owing to the comparable frequencies. In addition, the phase diagrams almost keep the same as the direction of ${\bf q}$ varies (${\phi\,=0^\circ}$ in Fig. 5(b) and ${\phi\,=30^\circ}$ in Fig. 5(f)).

The Coulomb excitations are greatly diversified by the stacking configuration and layer number.  Monolayer graphene, with the linear Dirac-cone structure, only exhibits the interband excitations in the absence of carrier doping. The 2D acoustic plasmon is absent, since this system is a zero-gap semiconductor with a zero DOS at ${E_F=0}$. However, the extra intraband excitations and acoustic plasmon could survive under the finite temperature and electron/hole doping, in which the latter experiences the serious Landau damping due to the interband e-h pairs at large momenta\cite{PRB34;979,RevModPhys83;1193,PRB87;235418}. As to a pristine bilayer AA  stacking, there are sufficient free carriers coming from the interlayer atomic interactions, creating two kinds of plasmons, namely, acoustic and optical modes\cite{PRB74;085406}. These two plasmon modes might be changed by the doping effect. On the other hand, the pristine bilayer AB stacking cannot induce acoustic and optical plasmons, mainly owing to very few free carriers associated with rather weak overlap in valence and conduction bands(b)). Both of them exist in the doped systems\cite{PRL81;4216}. There are rich and unique excitation behaviors in trilayer AB stacking. The e-h excitation boundaries, being defined by the distinct Fermi surfaces/the band-edge states, become more complicated. One acoustic and two optical plasmon modes are, respectively, related to the intraband and interband excitations of conduction electrons.

The high-resolution EELS could serve as the most powerful experimental technique to investigate the Coulomb excitations in emergent layered systems, such as, few-layer graphene, silicene, germanene, tinene and phosphorene. The EELS measurements on single- and few-layer graphenes have been used to confirm the plasmon modes, respectively, arising from the free carriers, all the $\pi$ electrons, and the $\pi+\sigma$ electrons. Specifically, the low-frequency acoustic plasmon (about below 1 eV) is identified to experience the interband Landau damping at larger momenta\cite{PRL83;161403,PRL80;113410}. The interband $\pi$ and $\pi$$+$$\sigma$ plasmons are observed at frequencies higher than 4.8 eV and 14.5 eV, in which their frequencies grows with the increase of layer number\cite{PRL83;161403,PRL80;113410}. However, the experimental identifications on the stacking-enriched electronic excitations are absent up to now. They are very useful in thoroughly understanding the diverse excitation phenomena closely related to the transferred (${\bf q\,,\omega}$)-phase diagrams. Furthermore, they provide the full information in examining the point of view that all the excitation behaviors are dominated by band structures.

The modified RPA is further developed to fully explore the doping effects on the Coulomb excitations of the AB-stacked trilayer graphene, in which the layer-decomposed bare response functions are introduced in the theoretical framework. The newly defined energy loss function is useful in understanding the diverse excitation phenomena and directly comparing with the experimental measurements. A pristine system only exhibits the obvious interband e-h excitations and cannot create any plasmon mode. Doping could dramatically alter the boundaries of the single-particle excitations, add the new excitation channels, and induce three kinds of plasmon modes. The first acoustic mode, the second and the third optical ones, respectively, originate from all the intraband $c_i\to\,c_i$ excitations,  the $c_2\to\,c_3$ and $c_1\to\,c_3$  interband  transitions. The last one could be observed only for $E_F$  crossing the highest conduction band. There exist the diverse ${({\bf q}\,,\omega\,)}$-phase diagrams, being sensitive to the doping carrier density, stacking configuration and layer number.

\par\noindent {\bf Acknowledgments}
This work was supported in part by the National Science Council of Taiwan, the Republic of China, under Grant Nos. NSC 105-2112-M-006 -002 -MY3 and National Research Foundation of Korea (201500000002559).

\newpage
\renewcommand{\baselinestretch}{0.2}

\newpage
\centerline {\textbf {Figure Captions}}

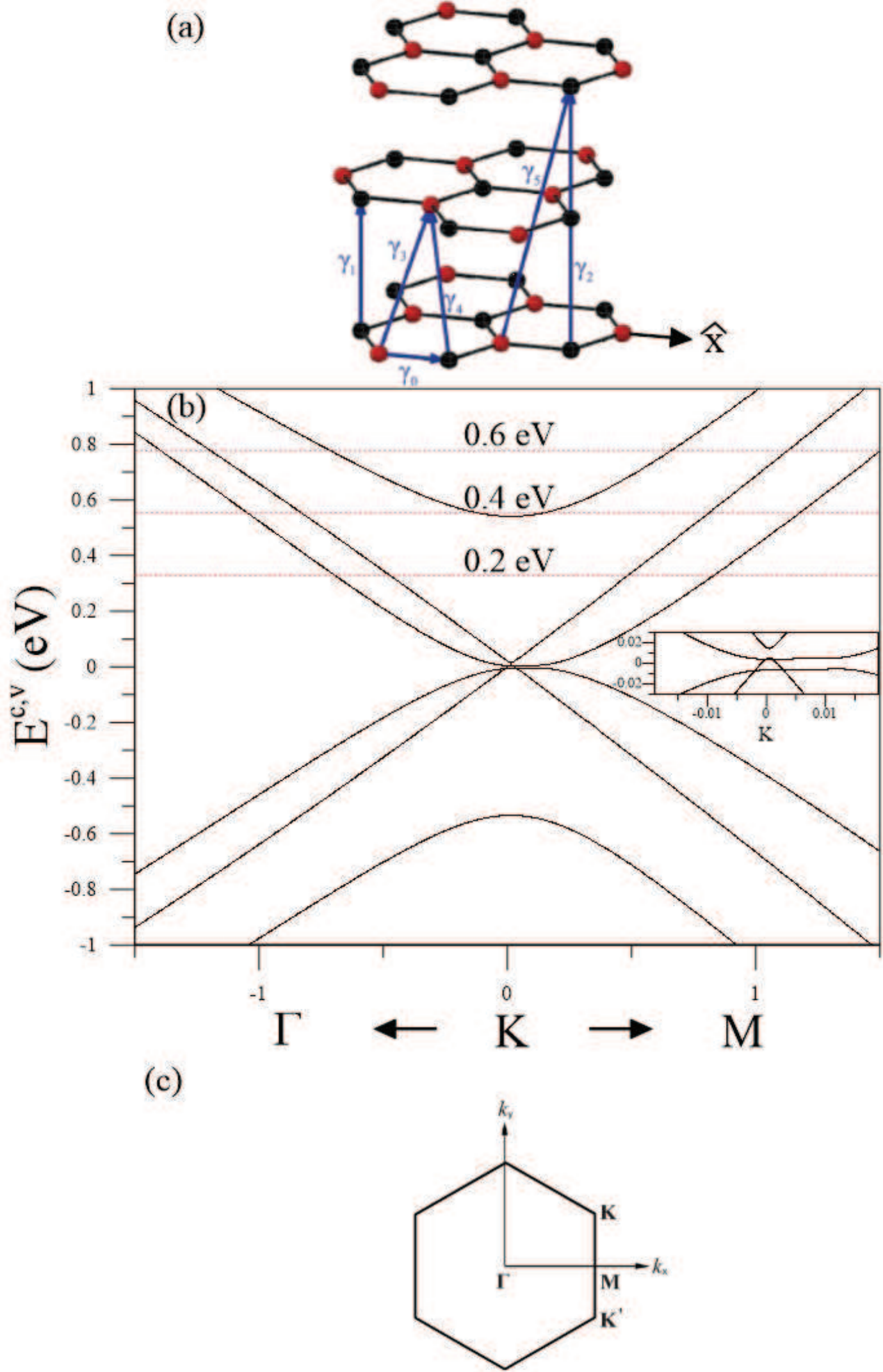
Fig. 1: (a) Geometric structure of AB-stacked trilayer graphene with the intralayer and interlayer atomic interactions, (b) a pristine band structure  along the high symmetry points, and (c) the first Brillouin zone.

Fig. 2: The independent four bare polarizations, (a) ${P_{11}}$, (b) ${P_{12}}$, (c) ${P_{13}}$ and (d) ${P_{22}}$,  for a pristine trilayer AB stacking at ${q=0.005}$ ${1/\AA}$ and ${\phi\,=0^\circ}$.

Fig. 3: Same plot as Fig. 1, but shown for an extrinsic system with ${E_F=0.2}$ eV.

Fig. 4: The energy loss functions at ${q=0.005}$ ${1/\AA}$ and ${\phi\,=0^\circ}$ under the distinct doping levels: (a) ${E_F=0}$, (b) ${0.2}$ eV, (c) ${0.4}$ eV and ${0.6}$ eV. For a ${E_F=0.2}$ system, they change with (e) ${q=0.005}$ ${1/\AA}$ $\&$ ${\phi\,=30^0}$, and (f) ${q=0.02}$ ${1/\AA}$ $\&$ ${\phi\,=0^0}$.

Fig. 5: The transferred momentum-frequency phase diagrams at ${\phi\,=0^\circ}$ for (a) ${E_F=0}$, (b) ${0.2}$ eV, (c) ${0.4}$ eV, (d) ${0.6}$ eV; (e) 0.8 eV. Also shown is that (f) under ${\phi\,=30^\circ}$ $\&$ ${E_F=0.2}$ eV.


\begin{thebibliography}{99}

\bibitem{JCP129;234709}
J. K. Lee, S. C. Lee, J. P. Ahn, S. C. Kim, J. I. B. Wilson, and P. John. J. Chem. Phys. $\mathbf{129}$, 234709 (2008).

\bibitem{JAP109;093523}
J. Borysiuk, J. Soltys, and J. Piechota, J. Appl. Phys. $\mathbf{109}$, 093523 (2011).

\bibitem{SurfSci610;53}
S. Hattendorf, A. Georgi, M. Liebmann, and M. Morgenstern, Surf. Sci. $\mathbf{610}$, 53 (2013).

\bibitem{PRL109;126801}
W. Yan, M. Liu, R. F. Dou, L. Meng, L. Feng, Z. D. Chu, Y. Zhang, Z. Liu, J. C. Nie, and L. He,
Phys. Rev. Lett. $\mathbf{109}$, 126801 (2012).

\bibitem{PRSLSA106;749}
J. D. Bernal, Proc. R. Soc. London, Ser. A $\mathbf{106}$, 749 (1924).


\bibitem{PRSLSA181;101}
H. Lipson, and A. R. Stokes,  Proc. R. Sot. London, Ser. A $\mathbf{181}$, 101 (1942).

\bibitem{PRL98;206802}
T. Ohta, A. Bostwick, J. L. McChesney, T. Seyller, K. Horn, and E.
Rotenberg, Phys. Rev. Lett. $\mathbf{98}$, 206802 (2007).

\bibitem{Science313;951}
T. Ohta, A. Bostwick, T. Seyller, K. Horn, and E. Rotenberg,
Science $\mathbf{313}$, 951 (2006).



\bibitem{RevModPhys83;1193}
M. O. Goerbig, Rev. Mod. Phys. $\mathbf{83}$, 1193 (2011).

\bibitem{PRB34;979}
Dielectric function and plasmon structure of stage-1 intercalated graphite,
Kenneth W. -K. Shung, Phys. Rev. B $\mathbf{34}$, 979 1986.

\bibitem{PRB87;235418}
S. Das Sarma and Qiuzi Li, Phys. Rev. B $\mathbf{87}$, 235418 (2013).

\bibitem{PRL81;4216}
S. Das Sarma and E. H. Hwang, Phys. Rev. Lett. $\mathbf{81}$, 4216 (1998).


\bibitem{ACSnano5;1026}
J. Y. Wu, S. C. Chen, O. Roslyak, G. Gumbs, and M. F. Lin, ACS Nano $\mathbf{5}$, 1026 (2011).

\bibitem{NanoLett16;6844}
T. Stauber, and H. Kohler, Nano Lett., $\mathbf{16}$, 6844 (2016).

\bibitem{PRB84;115420}
R. Roldan, and L. Brey, Phy. Rev. B $\mathbf{88}$, 115420 (2013).

\bibitem{JPSJ75;074716}
T. Ando, J. Phys. Soc. Jpn. $\mathbf{75}$, 074716 (2006).

\bibitem{NJP8;318}
B. Wunsch, T. Stauber, F. Sols, and F. Guinea, New J. Phys. $\mathbf{8}$, 318 (2006).

\bibitem{PRB74;085406}
J. H. Ho, C. L. Lu, C. C. Hwang, C. P. Chang, and M. F. Lin, Phys. Rev. B $\mathbf{74}$, 085406 (2006).



\bibitem{PRL105;016801}
S. J. Park, and R. E. Palmer, Phys. Rev. Lett. $\mathbf{105}$, 016801 (2010).

\bibitem{NanoLett14;3827}
F. J. Nelson, J.-C. Idrobo, J. D. Fite, Z. L. Mi$\check{s}$kovi$\acute{c}$, S. J. Pennycook, S. T. Pantelides, J. U. Lee, and A. C. Diebold, Nano Lett., $\mathbf{14}$, 3827 (2014).




\bibitem{PRL83;161403}
Shin, S. Y. et al. Observation of intrinsic intraband -plasmon excitation of a single-layer graphene. Phys. Rev. B $\mathbf{83}$, 161403 (2011).


\bibitem{PRL80;113410}
J. Lu, K. P. Loh, H. Huang, W. Chen, and A. T. S. Wee, Phys. Rev. B $\mathbf{80}$, 113410 (2009).


\bibitem{PRB61;7517}
D.Richards, Phys. Rev. B $\mathbf{61}$, 7517 (2000).


\bibitem{PRB43;4579}
J.-C. Charlier, X. Gonze, and J.-P. Michenaud, Phys. Rev. $\mathbf{43}$, 4579 (1991).



\bibitem{IOPBook;978}
C. Y. Lin, T. N. Do, Y. K. Huang, and M. F. Lin, IOP Concise Physics. San Raefel, CA, USA: Morgan $\&$ Claypool Publishers, 2017.

\end{thebibliography}
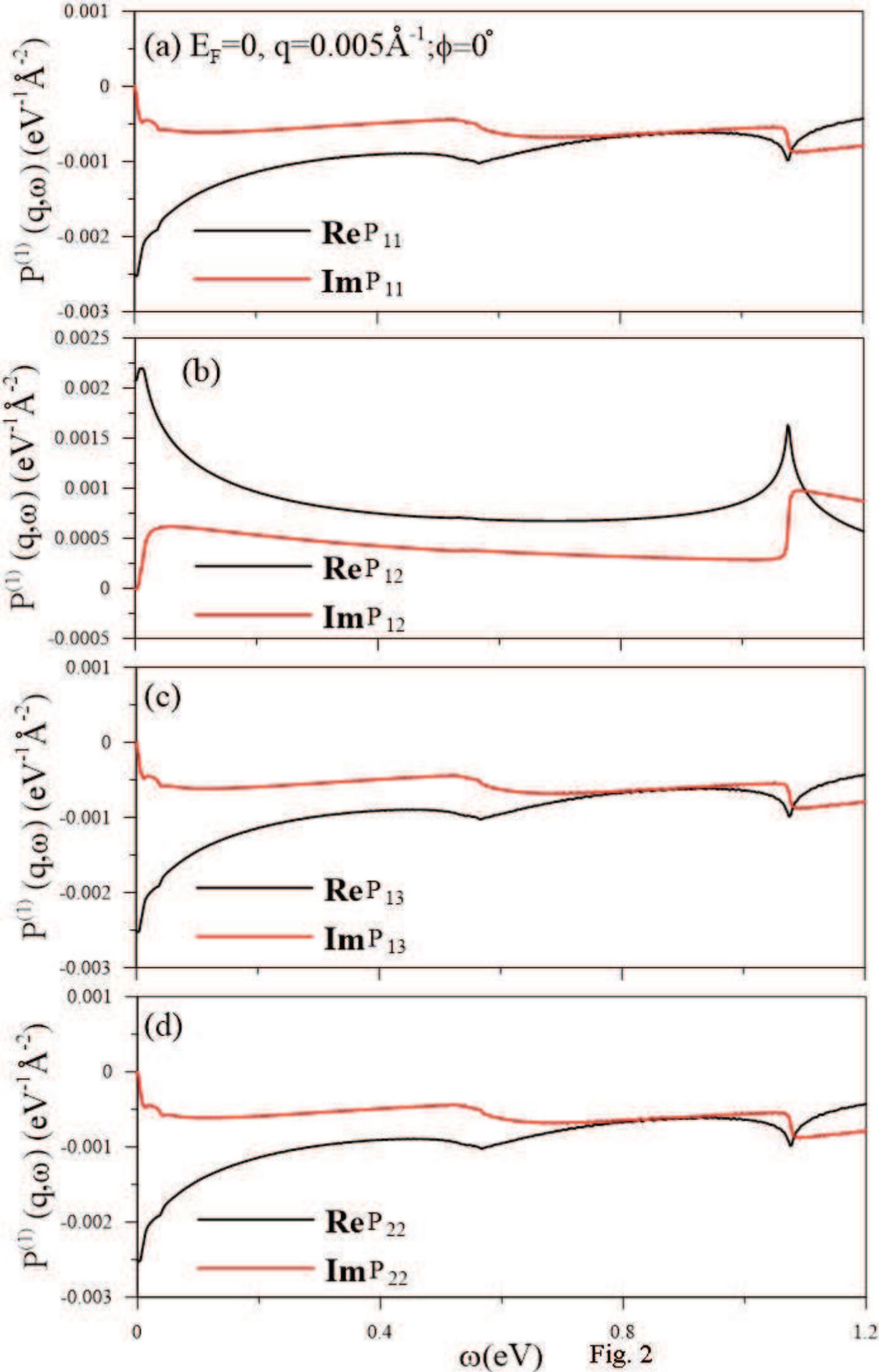
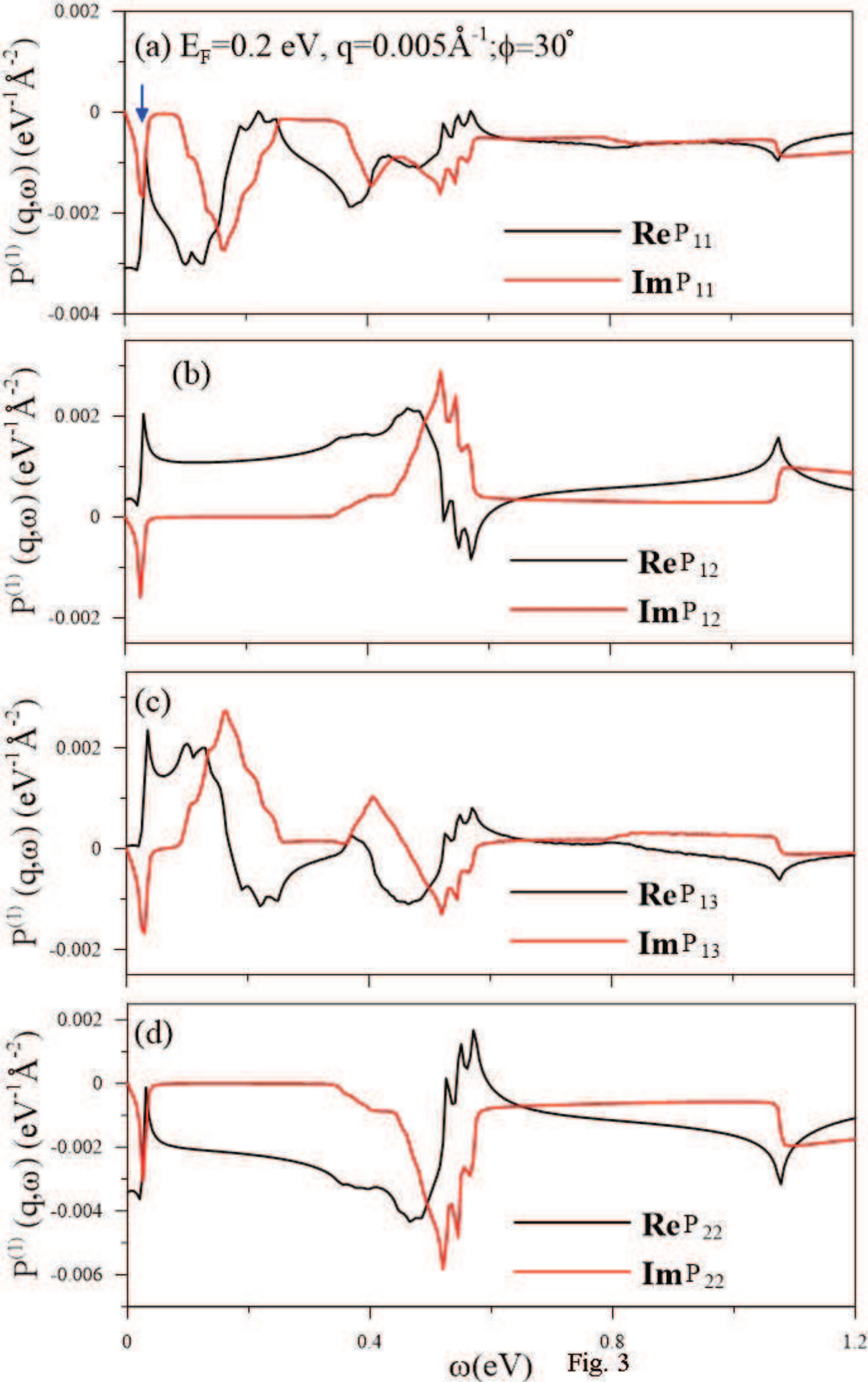
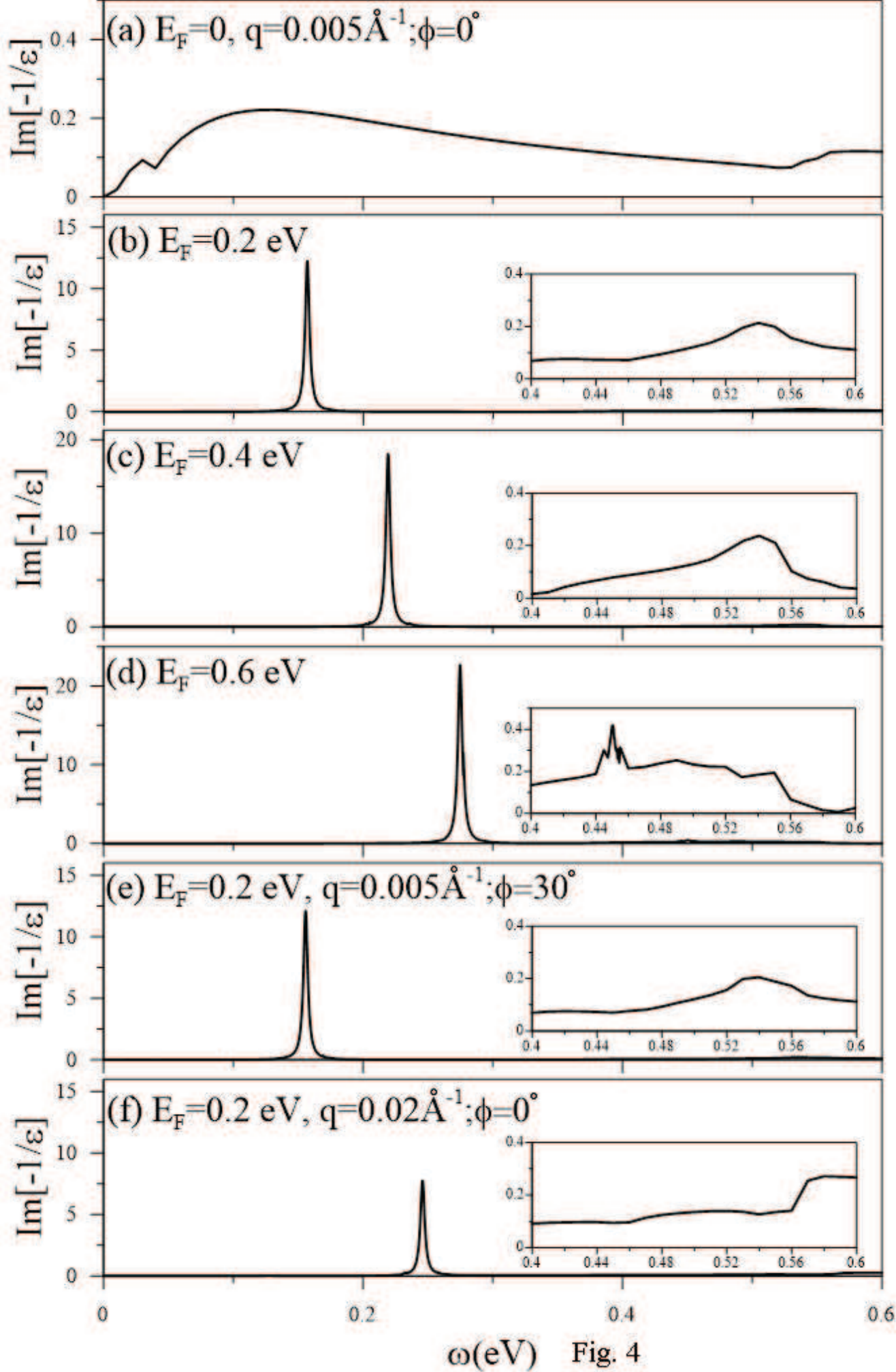
\end{document}